# Computing Shortest Non-Trivial Cycles on Orientable Surfaces of Bounded Genus in Almost Linear Time

Martin Kutz

Max-Planck-Institut für Informatik, Germany
mkutz@mpi-inf.mpg.de

**Abstract**

We present an algorithm that computes a shortest non-contractible and a shortest non-separating cycle on an orientable combinatorial surface of bounded genus in $O(n \log n)$ time, where $n$ denotes the complexity of the surface. This solves a central open problem in computational topology, improving upon the current-best $O(n^{3/2})$-time algorithm by Cabello and Mohar (ESA 2005). Our algorithm uses universal-cover constructions to find short cycles and makes extensive use of existing tools from the field.

## 1 Introduction

Decomposing a surface into topologically simple components is a basic technique in computational geometry. Many applications in computer graphics, like texture mapping, compression, and morphing, rely on decompositions of topological surfaces (see the overview in [3, Chap. 2.4] and the references in [9]).

Quite recently, the computational complexity of decomposition problems on combinatorial surfaces has gained growing attention. Erickson and Har-Peled [9] investigated the problem of cutting a topological surface efficiently into a disk and Colin de Verdière and Lazarus [6] showed how to decompose a given surface into parts of genus 0. Later works by Erickson and Whittlesey [10] and Colin de Verdière and Erickson [4] deal with the computation of short generators of the fundamental group and shortest paths and cycles in given homotopy classes. A central open question that arose in this context is:

**(P)** *How to compute a shortest non-trivial cycle on a topological surface efficiently?*

In other words, how to find the smallest cut that reduces the topological complexity of a given surface. There actually exist two very natural interpretations of the term "non-trivial," which are both studied in the aforementioned papers: non-contractible and non-separating cycles. The former are non-trivial with respect to homotopy, i.e., they are non-zero elements in the fundamental group of the surface, and the latter are non-trivial with respect to $\mathbb{Z}_2$-homology.

Erickson and Har-Peled [9] presented a first efficient, $O(n^2 \log n)$-time algorithm for this problem on orientable surfaces, where $n$ denotes the size of the surface. It forms a fundamental building block, and also the running-time bottleneck, for the path-shortening algorithm in [4].

A natural specialization of the general problem statement is to consider surfaces of fixed or bounded genus. This setting excludes the topological complexity from the problem and focuses on more essential algorithmic questions. The current-best algorithm for Problem (P) in the case of bounded-genus surfaces is due to Cabello and Mohar [2]. They show how to compute shortest non-contractible cycles in $O(n^{3/2})$ time and non-separating cycles in $O(n^{3/2} \log n)$ time, for orientable and non-orientable surfaces alike. (For the non-separating case, their algorithm actually yields subquadratic running time for any $g \in O(n^{1/3-\epsilon})$.) In the full version of their paper, they name the task of finding a near-linear-time solution to Problem (P) (possibly for surfaces of fixed topological type) as one of the most appealing open questions in the area.

The present paper solves this problem—on orientable surfaces—for both, the non-contractible and the non-separating case. We present an algorithm that computes a shortest non-trivial cycle on a combinatorial surface of bounded genus in $O(n \log n)$ time, where $n$ denotes the total description complexity of the surface.

Our algorithm builds heavily on previous techniques, mainly from [9, 10, 2, 7]. After a first introduction of the necessary notions and some formalism, we shall therefore revisit in a little greater detail the central algorithmic results from the field.

## 2 Topological Background and Formalism

We begin by fixing our formalism and notations for combinatorial surfaces and thereby also *briefly* recapitulate the most important underlying concepts from topology. This is not meant to, and clearly cannot be, an introduction to algebraic topology. So we only go into detail where rigorous definitions are indispensable. Beyond that, a fundamental familiarity of the reader with the basic concepts of homotopy (and also a little homology) is assumed. We refer to standard texts like [1] or [12] for a more thorough introduction to the matter.

A *combinatorial surface* is a 2-dimensional manifold $\mathcal{M}$ together with a graph $G$ embedded on $\mathcal{M}$ such that every face (i.e., every component of $\mathcal{M} \setminus G$) is a topological disk. For surfaces with boundary we further demand that every boundary component of $\mathcal{M}$ be a simple cycle in $G$. All surfaces considered in this paper are implicitly assumed to be connected.

For a representation of combinatorial surfaces, we follow the formalism of Colin de Verdière and Erickson [4]. We form the dual graph $G^*$ of $G$, which has a vertex $f^*$ for every face $f$ of $(\mathcal{M}, G)$, and an edge $e^*$ for every edge $e$ of $G$, connecting the two dual vertices $f_1^*$ and $f_2^*$ corresponding to the faces $f_1$ and $f_2$ that share the edge $e$. In case $\mathcal{M}$ has a boundary, $G^*$ also contains a vertex $\bar{e}^*$ for every boundary edge $e$ of $G$, which, in $G^*$, is connected by the edge $e^*$ to the dual $f^*$ of the $G$-face $f$ that contains $e$.

The pair $G, G^*$ (together with the correspondences of vertices, edges, and faces of the two graphs) contains all topological information about the combinatorial surface $(\mathcal{M}, G)$. As the *size* of the combinatorial surface we define the number of all vertices, edges, and faces of $G$ and $G^*$ together. In the following, we shall often omit the reference to either the graph $G$ or the underlying manifold $\mathcal{M}$, simply speaking of "the surface," implicitly assuming that we have a topological space $\mathcal{M}$ represented by a pair $G, G^*$ of graphs.

Usually, the surfaces under consideration in this paper will have no boundary. Only for technical reasons we might sometimes have to create boundaries by cutting along cycles. So unless stated otherwise, all surfaces are implicitly assumed to come without boundary.

All the algorithms considered in this paper work for weighted graphs as well as for the unit-distance setting. In the former, each edge $e$ of $G$ comes with a non-negative weight, while in the latter each edge has unit-weight 1. Edge weights in the dual graph will be defined when needed.

**Paths and cycles on a surface.** A *path* in a manifold $\mathcal{M}$ is a continuous map $\alpha$ from the unit interval $[0, 1]$ into $\mathcal{M}$. The images of 0 and 1 are considered the *start* and *end point* of the path, respectively. A *loop* (with *basepoint* $x$) in $\mathcal{M}$ is a path with $\alpha(0) = \alpha(1) = x$. A *cycle* is a continuous map $\gamma$ from the 1-sphere $S^1$ into $\mathcal{M}$. So loops and cycles are essentially the same thing, only that a loop comes with a distinguished basepoint. One usually identifies a path, loop, or cycle with its respective image in $\mathcal{M}$.

In a combinatorial surface, the above objects are embodied by graph-theoretic concepts. A *path* on a combinatorial surface $G$ is a (not necessarily simple) path in the graph $G$, i.e., a sequence of $G$-adjacent vertices. A *cycle* is a closed walk. A *simple* path or cycle contains no vertex more than once. The *length* or *weight* of a walk is the sum (with multiplicities) of all its edge weights. Note that this nomenclature differs from the graph-theoretic convention, where a "path" is usually already assumed to be simple and a "non-simple path" would be called a "walk." Here we chose to deviate from that tradition in order to avoid inconsistencies with standard topological notions.

**Contractible and separating cycles.** Two paths $\alpha$ and $\beta$ are *homotopic* if there exists a continuous map (a *homotopy*) transforming $\alpha$ into $\beta$, while keeping start and end points fixed. In particular, $\alpha$ and $\beta$ must have the same start and the same end point. Two loops are homotopic if they are homotopic as paths and two cycles are homotopic if one can be freely transformed into the other, i.e., no points are required to stay fixed.

The equivalence classes of all loops (with a fixed common basepoint $x$) form a group—the *fundamental group* $\pi_1(\mathcal{M})$ of $\mathcal{M}$—with concatenation of paths as group operation and the constant map as identity element. For a connected surface, the fundamental group is independent of the choice of the basepoint $x$. A loop that is homotopy equivalent to the constant loop (or cycle) is called *trivial* (w.r.t. homotopy) or *contractible*. We call a cycle or path *tight* if it is a shortest one in its homotopy class.

We distinguish two kinds of non-contractible cycles (or loops) on surfaces: *surface separating* and *non-separating cycles*. A simple cycle is *separating* if its removal from the surface decomposes the surface into more than one component. Every contractible cycle is separating. Technically, a separating cycle is trivial with respect to $\mathbb{Z}_2$-*homology*. Again, we refer to standard text books on topology for an introduction to homotopy and homology.

**The genus of a surface.** By now, our results seem to work only for orientable manifolds, so we assume from now on that *all surfaces under consideration are orientable*. The topological type (up to homeomorphism) of an orientable surface $\mathcal{M}$ is uniquely determined by its *genus* $g \geq 0$ and (in case $\mathcal{M}$ has a boundary) by the number $b$ of boundary components. Intuitively, $g$ counts the number of "holes" in the surface. For example, the torus has genus 1.

It is well-known that the fundamental group of a genus-$g$ surface without boundary is generated by $2g$ loops and moreover, one can choose the loops in such a way that each of them forms a simple non-separating cycle and any two of them intersect only in the common basepoint. If we now cut along all these loops, the result will be a topological disk. Erickson and Whittlesey [10] call such a collection of $2g$ almost-disjoint loops that generate $\pi_1(\mathcal{M})$ and whose removal leaves a disk, a *system of loops*. Figure 1 shows a system of loops on a genus-2 surface.

Technically, the existence of a system of loops is guaranteed only for topological spaces; on a combinatorial surface, there might simply not be enough space to route $2g$ loops without large overlaps. However, it turns out that these issues can be overcome. We shall address this point in more detail later.

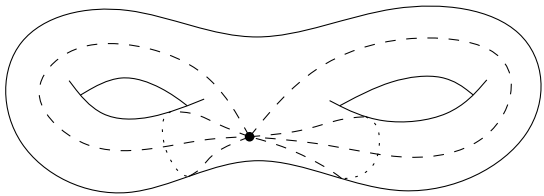

Figure 1: A system of loops on a 2-hole torus.

**The universal cover.** The *universal cover* of a connected topological space $\mathcal{M}$ is the (up to homeomorphism) uniquely determined simply-connected space $\tilde{\mathcal{M}}$ such that there exists a map $f$ from $\tilde{\mathcal{M}}$ onto $\mathcal{M}$ such that $f$ is locally a homeomorphism. The universal cover of any orientable surface of positive genus (and without boundary) is the infinite plane.

Fix a system of loops $\ell_1, \ldots, \ell_{2g}$ for $\mathcal{M}$. The universal cover $\tilde{\mathcal{M}}$ is tiled with infinitely many copies of $\mathcal{M} - \bigcup \ell_i$ (the surface $\mathcal{M}$ cut along the loops $\ell_i$), each such copy $D$ being called a *fundamental domain*. Two adjacent fundamental domains touch along a loop $\ell_i$. Every fundamental domain has each loop twice on its boundary, seeing it once from each side. If we orient each loop $\ell_i$ (arbitrarily) then its "left" and "right side" are well-defined and we denote their two occurrences on the boundary of a fundamental domain $D$ by $\ell_i^+$ and $\ell_i^-$, depending on whether $D$ lies to the left or right of the respective copy of $\ell$. See Figure 2

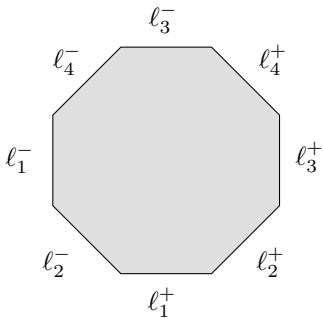

Figure 2: A fundamental domain of a 2-hole torus.

A path $\alpha$ in $\mathcal{M}$ can be *lifted* to a path in the covering space by choosing any inverse image of the starting point of $\alpha$ and then "unrolling" $\alpha$ into $\tilde{\mathcal{M}}$, making a transition between two adjacent fundamental domains whenever the path in $\mathcal{M}$ crosses a loop from the respective system of loops. This gives a procedure to decide whether two paths $\alpha$ and $\beta$ in $\mathcal{M}$ (with the same starting and end point) are homotopy-equivalent: After choosing the same starting point for a lift of $\alpha$ and $\beta$ into $\tilde{\mathcal{M}}$, we see that $\alpha$ and $\beta$ are equivalent if and only if their lifts also end on the same point in $\tilde{\mathcal{M}}$. In particular, a loop in $\mathcal{M}$ is contractible if and only if any, and thus all of its lifts are also loops.

## 3 Previous Algorithms

During the last few years, some powerful tools for the computation with paths and cycles on combinatorial surfaces have been developed. Several of those techniques will be used in our algorithm. Let us first briefly sketch the relevant results to introduce the reader to the methods. Later, when we present our new algorithm, we will address more technical details.

**Shortest non-trivial loops with given basepoint.** In [9], Erickson and Har-Peled described a fundamental method for finding a shortest non-contractible cycle through a given vertex $x$ on a combinatorial surface. Essentially, they just run Dijkstra's shortest-path algorithm, with starting point $x$. Whenever the wave front of equidistant vertices touches itself as it sweeps across the surface, the algorithm checks whether it does so in a trivial way, enclosing a topological disk on either side, or whether it found two homotopically different paths to that contact point, in which case a non-contractible loop has been found.

Altogether, a shortest non-contractible cycle containing a given vertex can be found in $O(n \log n)$ time (where $n$ denotes the complexity of the surface). With slight modifications, a shortest non-separating cycle can be found in the same time. The authors remark that this immediately gives an $O(n^2 \log n)$ algorithm for finding a shortest non-trivial cycle by simply running the above algorithm for once for each vertex of the surface.

**Constructing a whole system of loops.** The above algorithm for shortest-loop finding can be used to compute a whole system of loops. After computing a shortest non-separating loop $\ell_1$ through $x$, one can cut along this loop (implicitly duplicating every vertex on $\ell_1$) and then compute another "loop" from one copy of $x$ to the other in the new surface $\mathcal{M} - \ell_1$. Intuitively, the cut along $\ell_1$ forces the next loop, $\ell_2$, to lie in a different homotopy class than $\ell_1$. We may repeat this process of finding loops and cutting until $2g$ loops are found, which can easily be seen to generate the fundamental group of the surface. The resulting loops will in general not be disjoint but this technical detail is not too hard to resolve.

Erickson and Whittlesey [10] prove the remarkable fact that the greedy construction above always produces a *shortest system of loops* (i.e., a system of loops that minimizes the total length of all loops) for the given basepoint. They also present an alternative approach for computing a shortest system of loops through a given vertex in linear time for a surface of bounded genus. That construction uses Eppstein's tree-cotree decomposition, which we recapitulate next.

**The tree-cotree decomposition.** In [7], Eppstein presented a dynamic data structure for maintaining the embedding of a graph on a surface, which turned out to be a very powerful tool even in the static setting. For a given surface $\mathcal{M}$, two trees are computed: a rooted spanning tree $T$ (with root $x$) for the graph $G$ and a spanning tree $T^*$ for the dual graph $G^*$. The two trees must not intersect, in the sense that if an edge $e$ occurs in $T$ then its dual edge $e^*$ must not occur in $T^*$, and vice versa.

Let $X$ denote the set of all edges of $G$ that do not occur in $T$ and whose duals are not part of $T^*$. By Euler's formula, $|X| = 2g$. The triple $(T, T^*, X)$ is called a *tree-cotree decomposition* for $G$. Let $P(u)$ denote the unique path in $T$ from a vertex $u$ to the root $x$ of $T$. Each edge $e = \{u, v\} \in X$ induces a loop $\ell_e = P(u) \cup e \cup P(v)$ in $T$. Eppstein proves that these $2g$ loops form a homotopy basis and shows that for weighted graphs with all edge weights distinct, a minimum-spanning tree $T$ in $G$ and a maximum spanning tree $T^*$ in the dual $G^*$ are disjoint (in the above sense that at most one of $e$ and $e^*$ occurs in either tree) and can thus serve as tree and cotree, respectively, to form such a decomposition.

Erickson and Whittlesey [10] use the tree-cotree decomposition to obtain a shortest system of loops for a given basepoint $x$ very efficiently. First they compute a shortest-paths tree $T$, as above. Then they assign new weights $\omega(e)$ to the remaining edges $e \in E(G) \setminus E(T)$: for such an edge $e$ let $\omega(e)$ be the total length of the induced loop $\ell_e$. They then compute a *maximum* spanning tree $T^*$ in the dual $(G \setminus T)^*$, using as the weight of a dual edge $e^*$ the weight $\omega(e)$ of $e$, too. Again, an edge set $X$ of size $2g$ will be left and it can be shown that now $\{\ell_e \mid e \in X\}$ will actually be a shortest system of loops. (Again, the issue of overlapping loops is deferred.)

A great advantage of the tree-cotree decomposition over a repeated application of Erickson and Har-Peled's non-trivial-loop algorithm is that it runs in linear time on surfaces of bounded genus. More precisely, Erickson and Whittlesey observe that using an algorithm of Henzinger et al. [13], shortest-paths trees can be computed in linear time on a surface with genus $g \in O(n^{1-\epsilon})$. Note that "computing" here means providing the tree $T$ together with the set $X$ that induces the $2g$ loops. Simply outputting all resulting loops explicitly might require already $\Omega(gn)$ time.

**Routing through finite portions of the universal cover.** A standard technique for finding non-contractible cycles is to compute shortest paths in a finite portion of the universal cover. Assume we have a collection of cycles $\gamma_i$ whose removal decomposes our surface into a collection of topological disks. A system of loops, for example, has this property (producing exactly one disk). In the universal cover $\tilde{\mathcal{M}}$, each cycle $C_i$ appears as a collection of infinite paths. If we now have some bound on the number of times a potential path or cycle we are looking for can cross the cycles $C_i$, we know that a lift of the desired path or cycle can be found in a finite part of the universal cover.

Cabello and Mohar [2] use such an approach to find shortest non-contractible cycles and Colin de Verdière and Erickson [4] compute shortest paths and cycles within a given homotopy class by similar methods. While Cabello and Mohar work with systems of loops, Colin de Verdière and Erickson perform a so-called *tight octagonal decomposition* of the given surface such that the universal cover becomes a union of octagonal regions that are separated by shortest paths.

## 4 Our Algorithm for Shortest Non-Trivial Cycles

Our algorithm starts by computing a shortest system of loops as demonstrated in [10]. Precisely, we first compute a shortest-paths tree $T$ from some arbitrarily chosen basepoint $x$ and then find a maximum-spanning tree $T^*$ for the remaining dual graph with edge weights $\omega$ as discussed in the previous section. This takes $O(n)$ time if the genus $g$ of the surface is $O(n^{1-\epsilon})$ [10, Thm. 3.9].

Let us address the case of overlapping loops in a bit more detail. Cabello and Mohar [2], who also use the tree-cotree decomposition to find short cycles, describe a simple transformation of the underlying graph $G$ that turns all loops into simple cycles (still with basepoint $x$) that pairwise only intersect in $x$.

By construction, the two paths $P(u)$ and $P(v)$ that make up the loop $\ell_e$ for an edge $e = \{u, v\}$ from the set $X$, defined as above, intersect only on a path $\alpha$ starting from the basepoint $x$ until they follow different branches of the tree $T$. Cabello and Mohar duplicate all vertices along $\alpha$ (except $x$) introducing zero-weight edges between corresponding vertices so that afterwards $P(u)$ and $P(v)$ only touch at $x$, and $\ell_e$ thus becomes a simple loop. Similarly, they duplicate vertices wherever different loops $\ell_e$ and $\ell_{e'}$ for $e, e' \in X$ intersect. This whole process introduces no more than $O(gn)$ new edges and vertices and also requires $O(gn)$ time [2, Lem. 7].

One easily sees that the "split" loops $\ell_e$ ($e \in X$) now form a system of loops for the modified graph, which clearly maintained the topology of the original one. In fact, in our situation, this will still be a *shortest* system of loops because we have only introduced zero-weight edges into $G$. Formally this can be seen by running the whole construction of the tree-cotree decomposition from scratch for the modified graph. Taking care that none of the new edges is used for the tree $T$ (which can easily be guaranteed) we reproduce exactly the same system of loops that we obtained through the splitting process—which must now be a shortest system of loops by [10, Sec. 3.5].

**Controlling crossings.** A central ingredient for the shortest non-contractible-cycle algorithm in [2] is Lemma 7 from that paper, which guarantees the existence of a shortest non-contractible cycle that intersects the constructed system of loops not too often. We will employ that result for our algorithm, too. However, our formalism will deviate slightly from the one in [2] because we have to prepare for later needs and we also have to extend that lemma to non-separating cycles (which was implicitly mentioned in [2] already). Moreover, the respective lemma was stated with a very sketchy proof only. Therefore, we will present it here again, with full proof.

Intuitively it is clear what a "crossing" between two cycles on a surface should be. Yet, formally, we have to make precise whether and how two cycles are considered to "cross" if they share several consecutive edges. To this end, we think of every loop $\ell_i$ (and thus every of its edges) to have two "sides" and a path or cycle that uses such an edge $e$ from one of these loops is considered to lie by an infinitesimal margin to the left or right of this edge $e$. In the following, all cycles are implicitly assumed to come with fixed side informations for every single edge of them that lies on one of the loops $\ell_i$. Figure 3 shows a figurative drawing of this extra information.

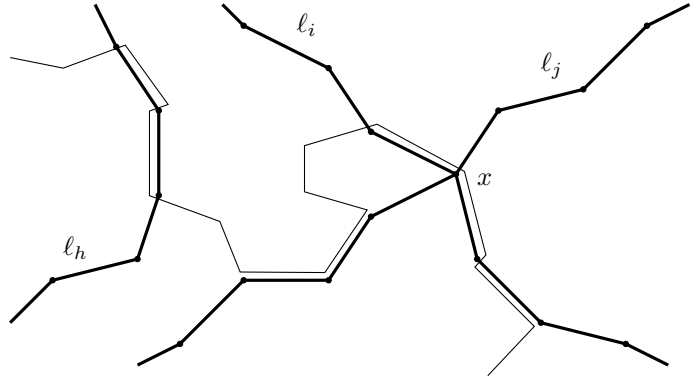

Figure 3: Every cycle contains side information for edges on the loops $\ell_i$.

Now, a *crossing* between a cycle $\gamma$ and a loop $\ell_i$ is a pair of consecutive edges $e, f$ in $\gamma$ such that the vertex of $\gamma$ at which $e$ and $f$ touch lies on $\ell_i$ and such that locally, $e$ and $f$ lie on different sides of $\ell_i$, taking the infinitesimal offsets into account, of course. The path in Figure 3, for example, crosses $\ell_i$ exactly twice, $\ell_j$ just once (at $x$), and $\ell_h$ three times.

There is a problem with cycles through the basepoint $x$. Since the loops $\ell_i$ only intersect at this point, this is the only vertex at which some cycle could intersect more than one loop at the same time. Such a situation would bring about some inconvenient technicalities, so we modify our graph $G$ a bit now in order to be able to ignore this special case in the future.

We subdivide every edge $e$ incident with $x$ into two edges and then connect the subdivision points in a cycle, in the order they occur on the surface. The weights of all the edges that are now incident with $x$ are set to some large number $L$ greater than all other edge weights in $G$ together, while the edges on the new cycle receive zero weight; the outer edges maintain the weight of the respective original edges. See Figure 4.

This modification obviously does not change the topology of the surface and it also does not change the distances between vertices—except those from and to $x$, of course. The advantage of the modified graph is that while the loops $\ell_1, \ldots, \ell_{2g}$ still all go through the origin, any shortest (non-trivial) cycle on the surface will now avoid the origin. This is easy to see. If some cycle went through $x$ in the original graph, we can reroute it on the zero-length cycle around $x$ now, to obtain a homotopically equivalent cycle of the same length as the original one. Conversely, any cycle in the modified graph maps to an equivalent one of same length in the old graph by shifting all vertices on the new cycle to $x$ and deleting repetitions. Our system of loops will still originate at $x$ and it will still be a shortest system of loops.

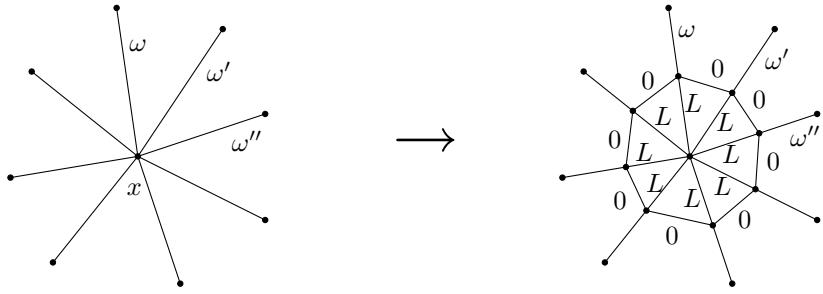

Figure 4: Blowing the basepoint up to a disk. The labels show edge weights.

The following lemma is a direct adaptation of Lemma 8 from [2].

**Lemma 1.** *Let $\ell_1, \ldots, \ell_{2g}$ be a system of loops for a surface $\mathcal{M}$ as computed above. Then there exists a shortest non-contractible and a shortest non-separating cycle on $\mathcal{M}$, each of them crossing any loop $\ell_i$ at most twice.*

*Proof.* We treat the non-contractible and the non-separating case simultaneously, speaking of "non-trivial" cycles to mean both variants and addressing either type explicitly when necessary. Note that we only need to consider simple cycles because any self-intersecting non-trivial cycle contains a proper subpath that forms a non-trivial cycle.

Amongst all shortest non-trivial cycles consider one, $\gamma$, that minimizes the total number of crossings with all $\ell_i$. If $\gamma$ intersects each of the loops at most twice then we are done. Otherwise, consider a loop $\ell_j$ that has at least three crossings with $\gamma$. Two of these crossings must lie on the same shortest path $P(a)$ in the tree $T$ between $x$ and one vertex $a$ of the edge that defined the loop $\ell_j$. Denote the $P(a)$-vertices at the two crossings of $\gamma$ with $\ell_j$ by $u$ and $v$, denote the subpath of $P(a)$ between $u$ and $v$ by $\alpha$, and call the two "halves" of $\gamma$ (separated by $u$ and $v$) $\gamma_1$ and $\gamma_2$. See Figure 5.

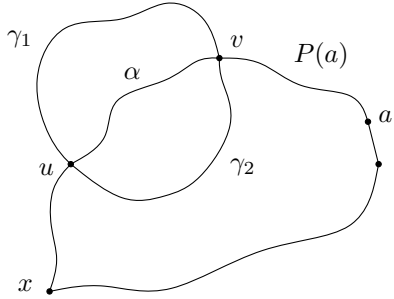

Figure 5: The situation from the proof of Lemma 1.

We have $|\alpha| \leq |\gamma_1|, |\gamma_2|$ because $\alpha$ as a subpath of $P(a)$ is a shortest path.

We claim that not both of the composite cycles $\alpha\gamma_1$ and $\alpha\gamma_2$ can be trivial. Assume for contradiction that they are. If they are both contractible then their composition $\gamma_1\alpha\alpha^{-1}\gamma_2^{-1}$, which yields the original cycle $\gamma$, must be contractible, too—a contradiction. In case the two cycles are separating, their sum (in $\mathbb{Z}_2$-homology) also gives $\gamma$, and thus a contradiction as well.

So we know that one of the two cycles $\alpha\gamma_1$ and $\alpha\gamma_2$ is non-trivial. Assume w.l.o.g. that the former is. This cycle is no longer than $\gamma$ and we can choose the side of the $\ell_j$ edges it uses in such a way that it has fewer crossings with $\ell_j$ than $\gamma$. If $\gamma_1$ connects to $\alpha$ at $u$ and $v$ from the same side of $P(a)$ then we place our new cycle completely to that side of $\ell_j$, thereby removing the two crossings at $u$ and $v$ and introducing no new ones. If $\gamma_1$ connects to $\alpha$ from different sides, we route the new cycle along an arbitrary side of $\ell_j$, which results in just one crossing, instead of two. In both situations, we do not create any new crossings with the other loops $\ell_i$ because the path $\alpha$ is completely disjoint from them. (At this point we make use of our modification of $G$ around $x$, which guarantees us that $\gamma$ does not touch $x$.) □

Lemma 1 tells us that when searching for a shortest non-trivial cycle, we need only consider such cycles that cross the system of loops a bounded number of times. Set in the universal cover, this means that we need only consider paths that pass through at most $4g + 1$ fundamental domains there. Naively applied, this approach would be quite inefficient, though.

Note that the above lemma used only the fact that $T$ was a shortest-paths tree, not that the cotree $T^*$ was chosen in any particular way. In the following we shall exploit the additional fact that the cotree $T^*$ was constructed in such a way that the resulting system of loops is a shortest one.

**Shortest cycles around cylinders.** In [4], Colin de Verdière and Erickson introduced a very useful lemma for tightening cycles. Using a planar $s$-$t$ cut algorithm by Frederickson [11], which improved upon techniques by Reif [14], the authors explain how to compute a shortest cycle on a cylinder $S^1 \times [0, 1]$ in almost linear time.

**Lemma 2 (Colin de Verdière & Erickson [4, Lem. 3.5(d)]).** *For a cylinder $S^1 \times [0, 1]$ of (combinatorial) size $n$, we can compute a tight cycle homotopic to the two boundaries $S^1 \times \{0\}$ and $S^1 \times \{1\}$ in $O(n \log n)$ time.*

Similar to [4], we will use this cylinder lemma as a tool to find short paths in the universal cover. However, we need a little more preparation before we can apply it to our situation.

**Lifting to unique fundamental domains.** Let us make the intuitive notion of path finding in the universal cover more precise. If we cut the graph $G$ along our system of loops, we obtain a plane graph $\hat{G}$ that has every loop vertex and edge twice on its boundary, except for $x$ which occurs exactly $4g$ times (see Figure 2 again). We envision the universal cover $\tilde{M}$ of $\mathcal{M}$ as represented by an infinite plane graph $\tilde{G}$, which is made up of infinitely many copies of fundamental domains $\hat{G}$, glued together along copies of the loops $\ell_i$.

Consider a lift $\tilde{\alpha}$ of some path or cycle $\alpha$ in $G$ that passes through several fundamental domains in $\tilde{G}$. Those edges of $\alpha$ that do not lie on one of the loops $\ell_i$ lie in a uniquely determined fundamental domain. For edges along the loops there are in principle always two possible fundamental domains they can be assigned to. However, the side information of $\alpha$ resolves this ambiguity; in the obvious way: an $\tilde{\alpha}$-edge on the left of a loop belongs to the fundamental domain to the left and an edge to the right belongs to the right-hand-side domain. This way, every lift $\tilde{\alpha}$ travels through a uniquely determined sequence of fundamental domains in the universal cover $\tilde{G}$. If $\alpha$ also avoids the basepoint $x$ (as is guaranteed with our graph) then any two fundamental domains in such a sequence are adjacent, i.e., they touch along the lift of some loop $\ell_j$.

**Definition 1.** We call a sequence $\mathcal{D} = (D_0, \ldots, D_k)$ of fundamental domains $D_i$ in $\tilde{G}$ a *meta path* if every pair $D_{i-1}, D_i$ is adjacent (i.e., touches along a lift of some loop $\ell_j$). Let $\alpha$ be any path or cycle in $G$. A meta path $\mathcal{D}$ is a *meta lift* of $\alpha$ if some lift of $\alpha$ travels through the fundamental domains in $\mathcal{D}$ in the respective order, i.e., it starts in $D_0$, then enters $D_1$, then $D_2$, and so on, until it reaches its end point in $D_k$.

**Forming cylinders.** Algorithmically, we want to use meta paths to find cycles with the help of Lemma 2. Therefore, we have to turn meta paths into cylinders.

The key idea is quite simple: identify the domains $D_0$ and $D_k$ of some meta path—in the right way. Figure 6 shows a meta path through the universal cover of a torus, for which this merging can be done without any problems. Note that in both drawings, $D_1$ borders on the top loop of $D_0$ and $D_3$ on the right loop of $D_4$. As we shall see, cycles around such a cylinder will produce non-trivial cycles on the underlying torus.

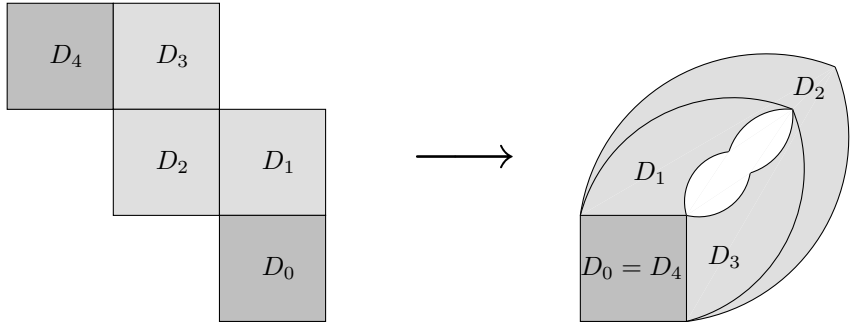

Figure 6: Merging the ends of a meta path.

Several details have to be taken into account for this gluing step, however. First of all, we have to deal with the possibility that some of the fundamental domains along $\mathcal{D}$ could be identical. In such a situation we always create a new copy of the fundamental domain so that $D_0, D_1, \ldots, D_k$ become all distinct. Technically, we do the following. For a given meta path $\mathcal{D} = (D_0, \ldots, D_k)$ we take pairwise disjoint copies $D'_0, \ldots, D'_k$ of the fundamental domains $D_0, \ldots, D_k$ and glue every pair $D'_{i-1}, D'_i$ along the boundary loops at which the corresponding fundamental domains $D_{i-1}, D_i$ touch, i.e., they get connected just like in the universal cover. For the constructed space, which is obviously a topological disk, we eventually identify the domains $D'_0$ and $D'_k$. Denote the resulting space by $\mathcal{D}^\circ$.

Is this space $\mathcal{D}^\circ$ always a cylinder? Unfortunately not. If some path $\alpha$ crosses the same loop $\ell_j$ twice, first from one side and then from the other, without crossing any other loop in between, then the respective steps in a meta path of $\alpha$ will be a sequence $D_{i-1}, D_i, D_{i+1}$ with $D_{i-1} = D_{i+1}$. In other words, a lift of $\alpha$ would enter some fundamental domain and then leave it through the same boundary $\ell_j^\pm$ of $D_i$. The resulting space $\mathcal{D}^\circ$ would thus contain the non-manifold configuration of Figure 7.

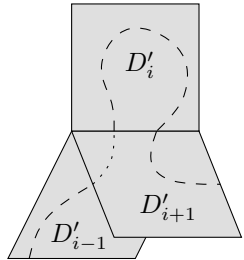

Figure 7: A non-manifold situation in the space $\mathcal{D}^\circ$.

However, we can show that there will always be a shortest non-trivial cycle that does not "curl" into a fundamental domain as the one in Figure 7 and will thus yield a space $\mathcal{D}^\circ$ that is a manifold.

**Definition 2.** We call a non-contractible cycle $\gamma$ in $G$ *curl-free* if for every meta lift $\mathcal{D} = (D_0, \ldots, D_k)$ of $\gamma$ we have $D_{i-1} \neq D_{i+1}$ for $1 \leq i < k$.

**Lemma 3.** *Amongst all tight cycles in the homotopy class of some given non-contractible cycle, the ones that cross the system of loops the least number of times are curl-free.*

*Proof.* Consider a tight non-contractible cycle $\gamma$ that is a shortest cycle in its homotopy class and which amongst all such tight cycles intersects the system of loops a minimum number of times. Assume for contradiction that for some lift $\tilde{\gamma}$ of $\gamma$ the corresponding meta lift $\mathcal{D} = (D_0, \ldots, D_k)$ has an index $i$ with $D_{i-1} = D_{i+1}$. Let $\varphi$ denote the subpath of $\tilde{\gamma}$ between the two crossing points $u$ and $v$ where $\tilde{\gamma}$ enters and leaves, respectively, the fundamental domain $D_i$. So $\phi$ "curls" into $D_i$, entering and leaving that fundamental domain through the same boundary loop, $\ell_j^+$ or $\ell_j^-$, say.

Replace $\varphi$ in $\tilde{\gamma}$ by the subpath of $\ell_j$ between $u$ and $v$, which cannot be longer than $\varphi$ because $\ell_j$ is from a shortest system of loops and the two paths are homotopic [10, Lem. 3.8]. Routing the new path segment on the $D_{i-1}$-side of $\ell_j$, we ensure that the resulting path $\gamma'$ now does not enter $D_i$ at all and thus has two fewer crossings with $\ell_j$ than $\tilde{\gamma}$. A contradiction to the minimality assumptions for $\gamma$ $\square$

Lemma 3 can be seen as an extension of Lemma 1. Now we know that some shortest non-trivial cycle in $G$ will not only cross each loop $\ell_i$ at most twice; its lift in $\tilde{G}$ will also always leave a fundamental domain through a different loop than the one through which it enters.

Note that this observation holds for non-*separating* cycles, too, not only for non-contractible ones, although Lemma 3 is only about the homotopy case. This is because we *first* apply Lemma 1 to learn that some shortest non-trivial cycle $\gamma$ intersects every loop at most twice and *then* conclude that in the homotopy class of $\gamma$ a shortest cycle with the minimum number of crossings must be curl-free, irrespective of whether $\gamma$ was separating or not.

Before we give a formal proof that for some shortest non-trivial cycle the space $\mathcal{D}^\circ$ will indeed be a cylinder, let us slightly modify that space to make the proof a little easier. Recall that by our modification around the basepoint $x$, no shortest non-trivial cycle will pass through $x$ anymore. So we always delete all copies of $x$, together with the whole disk of incident faces, from the space $\mathcal{D}^\circ$. This modification has no effect on the eventual algorithm but it makes the proof of the following lemma a bit easier.

**Lemma 4.** *There exists a shortest non-contractible (non-separating) cycle $\gamma$ with some meta lift $\mathcal{D} = (D_0, \ldots, D_k)$ with $2 \leq k \leq 4g$ such that the space $\mathcal{D}^\circ$ forms a cylinder.*

*Proof.* Lemma 1 guarantees the existence of a shortest non-trivial cycle in $G$ that crosses each loop $\ell_i$ at most twice and Lemma 3 tells us that if amongst all such cycles we take one, $\gamma$, that crosses the system of loops the least number of times then it will also be curl-free.

Consider a meta path $\mathcal{D} = (D_0, \ldots, D_k)$ of $\gamma$ By the crossing bound above we know that $k \leq 4g$ and a lift of any non-trivial cycle must pass through at least two different fundamental domains, hence $k \geq 2$. The curl-freeness of $\gamma$ guarantees that $\mathcal{D}^\circ$ is a manifold. There are only three possible topological types that could in principle result from the gluing process: a cylinder, a Möbius strip, or a disk. The orientability of $\mathcal{M}$ excludes a Möbius strip and the deletion of all copies of the basepoint implies that the resulting space cannot be simply connected. Hence, it must be a cylinder as claimed. $\square$

The next lemma shows that the cycles we want to compute in the meta cylinders $\mathcal{D}^\circ$ correspond to short cycles in $G$ as intended.

**Lemma 5.** *Let $\alpha$ be a curl-free non-contractible cycle in $G$ and let $\mathcal{D}$ be a meta lift of $\alpha$. Then the lift of $\alpha$ into $D$ forms a cycle $\beta$ in $\mathcal{D}^\circ$ homotopic to the two boundaries of $\mathcal{D}^\circ$ and of the same length as $\alpha$. Conversely, every non-contractible cycle $\beta$ in $\mathcal{D}^\circ$ homotopic to the boundaries of the cylinder corresponds to a cycle of the same length in $G$ and homotopic to $\alpha$.*

*Proof (sketch).* We can cut any cycle $\beta$ in $D^\circ$ at an arbitrary point in $D_0$ to obtain a path $\beta'$ from a point $b_0$ in $D_0$ to a point $b_k$ in $D_k$ through the universal cover. Since $b_0$ and $b_k$ correspond to the same point in $\mathcal{M}$, such a path $\beta'$ corresponds to a cycle in $\mathcal{M}$. Obviously, homotopic cycles in $D^\circ$ correspond to homotopic cycles in $\mathcal{M}$. □

The algorithmic realization of the above results is now straight-forward. List all different meta paths of length at most $4g$ in the universal cover $\tilde{G}$ of $G$. Form the space $\mathcal{D}^\circ$ from each of those meta paths and throw away all those spaces that are not cylinders. Use Lemma 2 to compute a shortest non-contractible cycle around each cylinder $\mathcal{D}^\circ$. This way we obtain a collection of cycles $\gamma$ in $G$. By Lemma 5 we know that one of these cycles must be a shortest non-trivial cycle of $G$.

Since a shortest non-trivial cycle must be simple, we throw away all cycles with self-intersections. For each of the remaining cycles, we can determine in linear time whether it is contractible or separating [9, Lem. 5.1].[1] This way we are bound to find a shortest non-contractible and a shortest non-separating cycle.

**Theorem 1.** *For an orientable combinatorial surface of bounded genus, a shortest non-contractible cycle and a shortest non-separating cycle can be found in $O(n \log n)$ time.*

*Proof.* We have already argued about the correctness of the above algorithm. The running-time bounds are easily checked. Our algorithm considers $O\big((4g)^{4g}\big)$ meta paths, which is a constant if the genus is bounded. A single cylinder computation costs $O(gn \log gn)$ time by Lemma 2, so the total running-time, for bounded $g$, is also $O(n \log n)$. The tests for contractible and separating cycles takes linear time per cycle. □

**Surfaces with Boundary.** We can use an idea from [9] to extend Theorem 1 to surfaces with boundary. In a preprocessing step, we glue a torus into each boundary component of $\mathcal{M}$. Precisely, for a boundary cycle $C$ with $r$ edges, we build a genus-1 surface with a length-$r$ boundary and attach it along $C$. All edges inside the new surface parts receive a common weight of some number $L$ larger than the sum of all edge weights in the original graph $G$. The whole process, for all boundary cycles together, will increase the combinatorial size of $\mathcal{M}$ by a linear factor, only. It is easy to see that a shortest non-trivial cycle on the modified surface will also be a shortest such cycle on the original surface and vice versa.

**An application: tightening paths and cycles.** In [4], Colin de Verdière and Erickson present algorithms to find shortest paths and cycles in given homotopy classes of an orientable surface. After precomputing a special decomposition of the surface into octagonal slabs in $O(n^2 \log n)$ time, they can find a shortest path homotopic to a given path of complexity (combinatorial length) $k$ in $O(gnk)$ time and a shortest cycle homotopic to a given cycle of complexity $k$ in $O(gnk \log(gnk))$ time.

We observe here that Theorem 1 can be used to speed up the precomputation of their algorithm from $O(n^2 \log n)$ to $O(n \log n)$, so that the running time of this step is now dominated by the later path computation and can thus be ignored.

---
[1] One could actually just read the relevant information directly off the meta paths but we prefer to avoid the necessary group theoretic argument here.

# 5  Conclusion

We have shown how to compute shortest non-contractible and shortest non-separating cycles on orientable surfaces of bounded genus in $O(n \log n)$ time. A question that naturally arises from this result is whether it can be translated to the case of non-orientable surfaces. Another, possibly more challenging problem would be to bring down the exponential dependence of the running-time on the genus of the surface. Maybe a clever inspection of the universal cover might yield a polynomial dependence and thus generalize the result to useful statements for surfaces of unbounded genus.

# Acknowledgements

This work was initiated during this year's Max-Planck ADFOCS summer school at Saarücken. Thanks to Jeff Erickson for showing me the problem and for discussing early ideas. I also thank Carsten Schultz for helpful discussions during the final stage of this work.